# How to adequately describe full range intercalation – a two-sided approach

Yue Zhu[1] and Joachim Maier[1]*

[1]Max Planck Institute for Solid State Research, Heisenbergstraße 1, 70569 Stuttgart, Germany

*office-maier@fkf.mpg.de

Understanding of the equilibrium cell voltage of an intercalation electrode as a function of the storage degree, is one of the key problems in battery research. Typically neutral lattice gas models are applied, but they may, for lithium ion batteries, at best satisfactorily interpret the situation very close to the potential of pure lithium [1]. For neutral lithium there is hardly any problem to fill the available sites, provided the shrinking number of empty sites is taken account of. However, under almost all circumstances lithium is, at least partly, dissociated. In fact, the dissociation into a well-accommodated lithium ion and a well-accommodated electron is the reason for high voltages. Such cases can be tackled by defect chemistry which treats the thermodynamics of incorporated ions and electrons in terms of formation/annihilation of the point defects being the relevant charge carriers.

Note that storage of lithium, i.e. incorporation of excess lithium in the lattice means occupation of interstitial sites or annihilation of vacancies as regards the ionic part, and formation of excess electrons or annihilation of electron holes as regards the electronic part. If we first concentrate on the incorporation into interstitial sites accompanied by the formation of excess electrons (conduction electrons), we can, using Kröger-Vink notation, write

$$Li + V_i \rightleftharpoons Li_i^· + e'  \qquad (1)$$

In Equation 1 $Li_i^·$ is the Li-ion incorporated in available interstitial positions ($V_i$) and $e'$ the excess electron simultaneously introduced. The difference of this consideration from the neutral model is most obvious in the Li activity $a_{Li}$. At this stage it is important to emphasize that the Li activity is not an abstract parameter, rather $\ln a_{Li}$ is linearly related to the Li chemical potential $\mu_{Li}$ and hence to the negative open circuit cell voltage. In other words, the concentration-activity relation provides the equilibrium charge/discharge curve. For very small Li contents, this parameter is given by the total Li concentration ($a_{Li} = x_{Li}$)[1] in the fully neutral case, while in the fully dissociated case the mass action law for Equation 1 demands that $a_{Li} = x_{Li}^2$. Again, this discrepancy is not a formal point; rather it changes the voltage-charge relation substantially and shows the necessity of considering non-idealities (interactions and saturation) for ionic and electronic species separately plus their mutual interaction.

Taking account of both dissociated and neutral species is easily possible via association concepts which take care of ion-electron interaction in a simple way according to

$$Li_i^· + e' \rightleftharpoons Li_i^x \qquad (2)$$

---

[1] Within this paper we define x and [defect] as molar ratios, e.g., relative to the number of moles of FePO$_4$ in LiFePO$_4$. The x should not be confused with the superscript in $Li_i^x$, which denotes the relative charge of zero.

where $Li_i^x$ is short for the neutral interstitial species.

With the electroneutrality equation,

$$[Li_i^·] \simeq [e'] \qquad (3)$$

it follows immediately from the mass action laws for dilute defect concentrations that $[Li_i^·] = [e'] \propto a_{Li}^{1/2}$ and $[Li_i^x] \propto a_{Li}$. As a consequence, $[Li_i^x]/[Li_i^·] \propto a_{Li}^{1/2}$ in agreement with the above statement that the higher $a_{Li}$, the more the Li interstitials tend to be neutral. While this configurational entropy effect is specific to the interstitial Li (see below), there is however an even stronger and very general contribution to the positive activity dependence which is energetic in nature, and applies if different compounds are to be compared. When this is the case, the mass action constant for Equation 1 and hence the proportionality constant (in front of $a_{Li}^{1/2}$) changes. It is determined by the exponential of the free energy of forming neutral defects and thus essentially by the energy level of the electrons [2]. In cathode materials the electron is easily accommodated by the transition metal centers while in anode materials the electronic energy is rather high. Consequently, when however the Li activity is sufficiently low (voltage vs. Li sufficiently high) Li will be fully dissociated and ions and electrons have to be considered separately. This is equivalent to the necessity of applying ionic and electronic statistics including interaction effects. As far as the electronic part is concerned, two extreme cases are possible: the strictly localized picture (certain transition metal ions change valence) and the highly delocalized picture (filling a band). In the first case one might treat the electrons (small polarons) similarly as the ions; in the latter, the rigid-band model is usually considered to be a good first approximation [3], and for small Li contents the parabolic band approximation further simplifies the situation. For large Li contents, these parabolic band approximations are no longer valid, but more importantly electron-electron interactions lead to severe deviations from the rigid-band model. As far as ion-electron interaction is concerned, the sheer description in terms of short-range interactions (see Equation 2) is insufficient and Debye-Hückel [4], effective defect-lattice [5] or more sophisticated long-range corrections [6-8] have to be considered. Certainly all these non-idealities are hard to treat.

**(1) Two-sided approach and thermodynamic interpolation procedure for $Li_xFePO_4$**

Here we will show using the example of $Li_xFePO_4$ that the problem is significantly simplified if one approaches the situation from both ends of the storage range (x=0 and x=1). Interpolation into the middle using the thermodynamic criterion for chemical phase stability may even make the exact treatment of non-idealities obsolete if only a phenomenological description of the theoretical charge-discharge curve is intended. In the following we will set out this enormously helpful concept for $Li_xFePO_4$, but we will – in the next section - also clarify how such non-idealities can be treated, which leads to mechanistic insights.

$LiFePO_4$ is one of the most relevant Li-ion battery electrodes, as it shows very good performance data and is considered to replace the present high-voltage electrodes owing to environmental benignity and abundance of the constituting elements [9]. For us it is a perfect model material. In the macroscopic state $FePO_4$ can only take up 5% Li, then transforms to $LiFePO_4$ with a maximum deficiency of 11% [10]. Such end-member situations can be handled rather straightforwardly in the intrinsic as well as extrinsic (doped) case by defect chemistry [11,12]. In the nano-crystalline case though there is full solubility over the entire range [13], and it is our aim to describe such a complete equilibrium charge/discharge curve[2].

Before presenting our approach from both ends, the deficiency of the neutral model is illustrated first. The litmus test is consideration of the limiting situation $x_{Li} \to 0$ where all the non-idealities should asymptotically vanish. Figure 1a plots the experimental voltage (V) vs. $\ln x_{Li}$ in the initial region of the discharge curve. The above equations indicate that, for dilute Li concentrations (small $x_{Li}$), the limiting slope of $-\frac{V \cdot F}{RT}$ vs. $\ln x_{Li}$ is 1 for the neutral model but 2 for the case of complete dissociation (Appendix 1). Clearly 2 is identified as limiting slope for FePO$_4$.

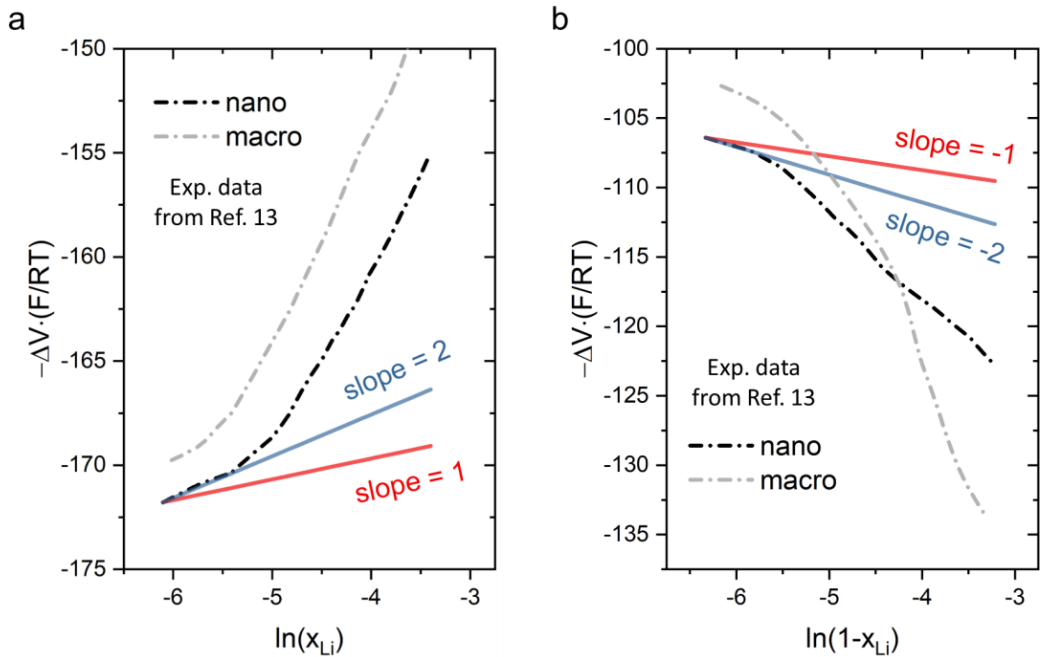

Fig. 1: Comparison of the neutral lattice-gas model and the complete dissociated model with cell voltage values for nano- and macro-crystalline iron phosphate.

---

[2] We refer to the very systematic and comprehensive experimental data given in Ref. 13.

As the above treatment has clearly shown the defect model treating the Li as mostly dissociated to be the right approach, we can now turn to the defect formulation at the other end (very large Li contents) close to the stoichiometric LiFePO$_4$. Unlike for FePO$_4$ where we designated the slight Li-excess in terms of interstitial defects and used Equation 1, we describe the slight Li-deficiency in the almost completely filled iron phosphate (i.e. LiFePO$_4$) in terms of vacancies and write

$$\text{Li} + V'_{\text{Li}} + h^{\cdot} \rightleftharpoons \text{Li}^x_{\text{Li}} \qquad (4)$$

The situation is symmetrical to the above, and we just have to consider vacancies ($V'_{\text{Li}}$) and holes ($h^{\cdot}$) instead of interstitials ($\text{Li}^{\cdot}_i$) and electrons ($e'$). The formation of neutral defects is then to be described by

$$V'_{\text{Li}} + h^{\cdot} \rightleftharpoons V^x_{\text{Li}} \qquad (5)$$

Symmetrically to the aforementioned treatment for very small Li contents, we now have for dilute vacancy conditions $[V'_{\text{Li}}] = [h^{\cdot}] \propto a_{\text{Li}}^{-1/2}$, $[V^x_{\text{Li}}] \propto a_{\text{Li}}^{-1}$, and $[V^x_{\text{Li}}]/[V'_{\text{Li}}] \propto a_{\text{Li}}^{-1/2}$. (Note that within the LiFePO$_4$ vicinity the association degree now decreases with the Li-activity owing to configurational entropy. This does however not contradict to the statement that due to energetic reasons the more anodic the situation, the higher the tendency to form neutral Li. This latter statement namely refers to variation of compounds.) This limiting situation predicted by the dissociated model is again confirmed by Figure 1b.

It is revealing to compare the situation with the results for the macro-crystalline $FePO_4$ and $LiFePO_4$. As Figure 1a shows, also here the quadratic relation is fulfilled as limiting law. Slope as well as absolute voltage values are different and the differences are much larger for the $LiFePO_4$ end (Figure 1b). This is reflected by different mass action constants (defect energies) owing to size effects, one being the capillary pressure term (cf. Fig. 22 in Ref. 2). Also the lower delocalization in nano-crystals may influence the slope in the same way. Focusing only on the nano-crystalline case where we have no miscibility gap, it is important to note that the slopes rapidly deviate toward values larger than 2, indicating existence of non-idealities even at small Li contents. But let us come back to the nano-crystalline material where we have full solubility.

Before we start detailing the treatment of such non-idealities, let us make the following key point: already the simple ideal mass action law description without interactions suffices to satisfactorily reproduce the entire charge discharge curve provided we proceed as follows: (i) We use the two-sided approach. (ii) We consider filling statistics (as outlined below). This is, as we will show, essentially important for the delocalized electrons. (iii) When for larger non-stoichiometries severe deviations occur from the experimental curve, we leave our interpretative approach and interpolate between the two rather dilute regions (see Figure 2). As interpolation tool we use the strict thermodynamic criterion known from chemical thermodynamics that for a phase to be stable the slope must be strictly negative, in other words, the chemical capacitance must be positive. Such criteria are well-treated in the chemical thermodynamic literature [14,15]. In the language used in Ref. 15 thermal capacity, mechanical capacity and chemical capacities need to be positive as otherwise any perturbation leads to a run-away of the relevant thermodynamic potential (Appendix 2). In other words, $\partial\mu_{Li}/\partial x_{Li}$ is required to be positive and hence the slope in the theoretical charge-discharge curve must be negative. Moreover we can safely smoothen out the joinings as the chemical potential and hence the chemical capacitance must be smooth functions. Therefore there cannot be irregularities and one can easily interpolate between the two curves of the end-members (FePO$_4$ and LiFePO$_4$) as shown in Figure 2.

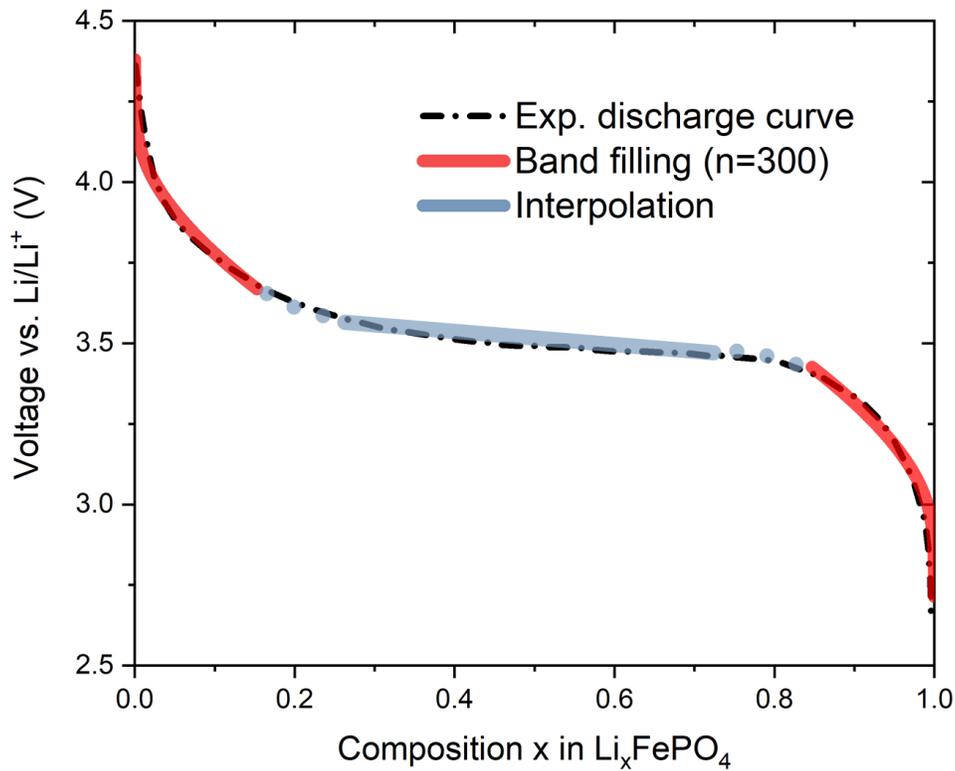

Fig. 2: Fitting of the defect chemical model incorporating band filling (see below) and using the two-sided approach (middle part deviating from the experimental curve is replaced by a simple linear interpolation whereby the joinings are smoothened out).

Figure 2 shows how well this pragmatic approach functions even though no interactions have been included. In spite of the success of this procedure, we will – in the following – more specifically tackle non-idealities as their inclusion is indispensable if one aims at a mechanistic understanding of the full-range curve. We will still use our two-sided approach, since it is adequate from a microscopic point of view and since it simplifies the problem enormously.

Let us now describe deviations from the dilute situation by first considering the filling statistics and then the interactions. Owing to symmetry it suffices to refer to the FePO$_4$ side (i.e. x<0.5).

**(2) Filling statistics**

The most obvious and necessary deviation from ideality is taking account of the finite number of sites (ions) and states (electrons). The simplest case is the ionic case as the sites all have the same single (free) energy. Then the simple saturation statistics leads to replacing x by x/(1-x). This is a Fermi-Dirac type of statistics for a delta-function-like density of states. In fact, if the regular states ($V_i$) are explicitly considered in the mass action law, it suffices to apply Boltzmann statistics (as $x_{Li} = [Li_i]$ and $1 - x_{Li} = [V_i]$). This correction is essentially perceived for large x and is felt within both halves of the description, as Figure 3a shows. Even though the correction is small, it points in the right direction. The same formalism could be used for strictly localized electrons, as then the two valence states of Fe enter the mass action laws (Appendix 3). If however the electrons are delocalized, the energy soon increases on filling. Then the Fermi-Dirac integral of the density of states has to be involved. For a rigid band and a parabolic band (which might be a good approximation for small non-stoichiometries), analytic approximations are available such as the one recommended by Nilsson [16], which is the one we will apply. A fitting parameter n (related to a material's physical properties) is available to tune the resulting curve and its effect, which is solely a result of filling available states of electrons, is huge on the chemical potential of Li hence the cell voltage (Appendix 4). Figure 3b indicates that a realistic fitting parameter of n=300 leads to a rather good agreement for non-stoichiometries less than 0.2.

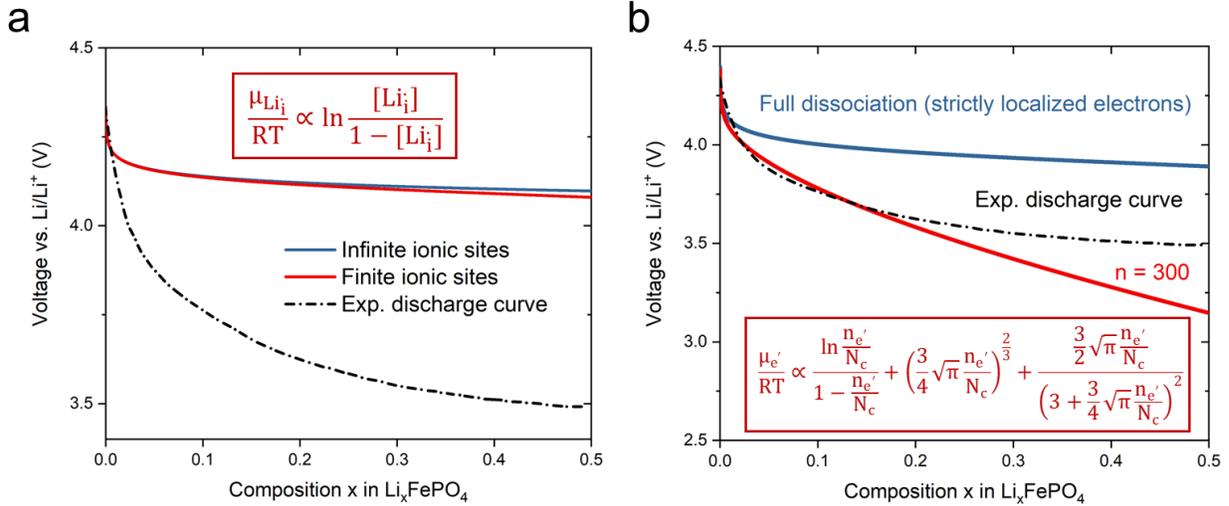

Fig. 3: Voltage change affected by (a) finite number of sites of ions and (b) finite states of electrons.

As all the other corrections described or to be described have a small effect or an effect that worsen even larger the agreement with the experiment, we conclude that band filling effects are the most important ones. This refers to the range of small non-stoichiometries. For large values, we recognize from Figure 3b that the band filling correction leads to an over-shooting pointing towards the necessity of including interactions. Please note that in the pragmatic approach which has been described in the previous section and has resulted in Figure 2, saturation effects (unlike the interaction effects to be described in the next section) have been taken into account. This is worth mentioning for the electrons (n=300), while the ionic effect is marginal at these concentrations.

## 3. Ion-electron interactions

According to the Bjerrum conception, interactions between differently charged species can be divided in short-range effects and long-range effects with the dividing line being determined by the Bjerrum length [17]. The short-range interaction can be well described by association equilibria as exemplified in the beginning (Equation 2 and 5). Here another fitting parameter $K_a$ (the association mass action constant) is at our disposal and its meaning is self-explanatory. Figures 4a,b show the effect of partial association when ionic or electronic saturation is included. Obviously, the contribution of partial association and electronic saturation brings the ΔV-curve close to the experimental curve.

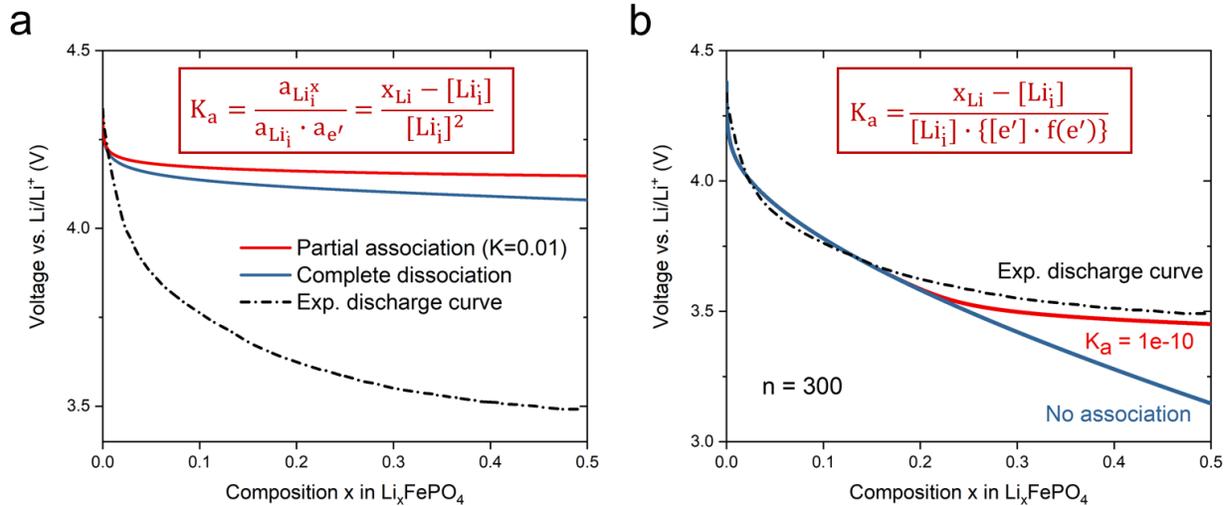

Fig. 4: (a) Voltage change caused by partial association ($K_a$=0.01) including ionic saturation (see Figure 3a). (b) Effect of the association including electronic saturation (see Figure 3b). Equations of the insert are explained in Appendix 5.

Figure 5 reveals how well the combination of band-filling and association describes the entire curve if the analogous procedure has been performed for the LiFePO$_4$ side (Appendix 6). As however some of the interactions (in particular electron-electron interactions) cannot be small, the marvelous agreement is certainly somewhat fortuitous. This may also be the reason why the n-values for the two end-members are found to be quite different.

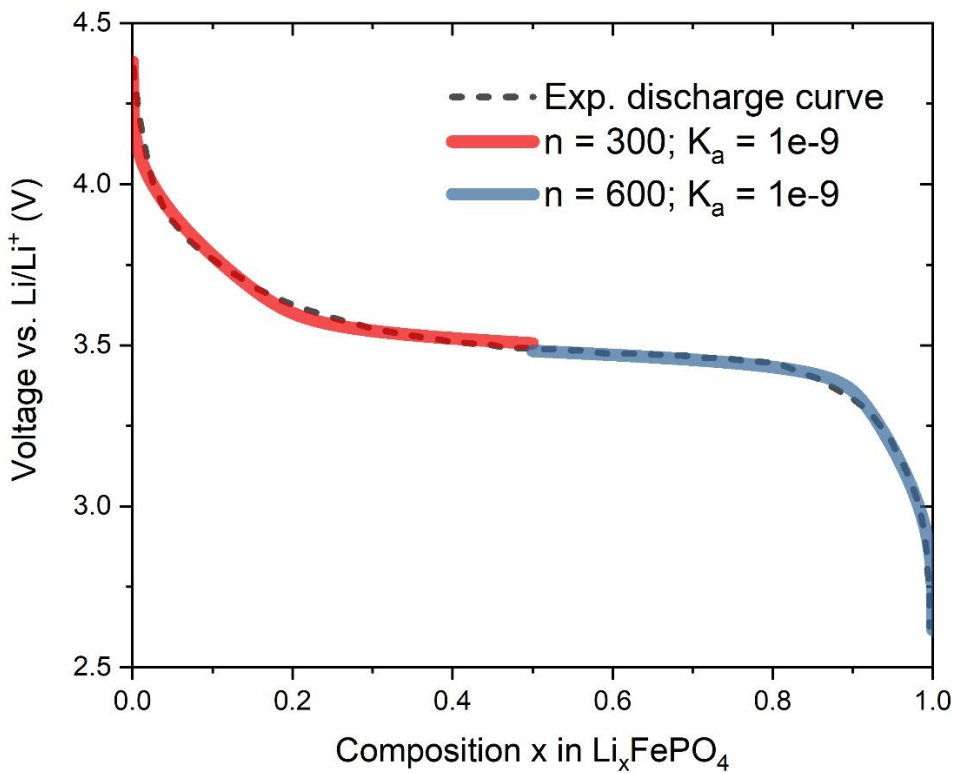

Fig. 5: Overall modeled discharge curve combining band filling and association from both sides.

Long-range interactions between ions and electrons are certainly more complicated to describe. While association lowers the concentration of charge carriers, long-range attractive Coulomb interaction leads to a concentration increase. The simplest approach is Debye-Hückel, where the effective excess may increase with the square-root of the concentration. This approach soon loses its validity while more sophisticated models [6-8] are hard to handle. The effective defect-lattice model [5] which leads to a cube root dependence does often good service. However, these corrections are not tailored for delocalized electrons but should be helpful in a first approximation. All these long-range models lead to a "wrong" correction when experimental values are considered and seem however not very important in our context. This does not mean that they are irrelevant as there might be compensation effects (cf. (5)).

**(4) Ion-ion interactions**

In our context ion-ion interaction refer to interactions between alike charges. As the long range interaction between opposite charges (Debye-Hückel or defect-lattice) includes also the interaction with the alike charges, consideration of ion-ion interaction alone is obsolete. As discussed above, the overall effect leads to a "wrong correction".

**(5) Electron-electron interactions**

The remaining interaction is electron-electron interactions that are not included in the consideration in (3), i.e. quantum mechanical interaction. Such eon-eon interactions lead to deviations from the rigid band model. For the latter, the introduction of a perceptible Hubbard energy is the usually used procedure [18]. In fact, none of the $LixFePO4$ compositions have been shown to exhibit high electronic conductivities [19] and individual band structure calculations seem to be required for any x. Obviously covering the full range with simple models appears hopeless. To reiterate, the seemingly perfect fitting shown in Figure 5 may be partly coincidental considering the fact that Li$_x$FePO$_4$ does not follow a rigid band behavior [A. Yaresko, personal communication].

Conclusions

We treated the full incorporation thermodynamics using nano-crystalline Li$_x$FePO$_4$ as a master example as here the full range from x=0 to 1 is experimentally accessible. The following points are most important: The treatment in terms of the neutral lattice-gas model is incorrect, rather ions and electrons need to be considered separately. One needs to invoke point defect chemistry which gives a satisfactory agreement all over the range if we treat the problem from the two-sides: from the FePO$_4$ side where Li-ion is incorporated interstitially and from the LiFePO$_4$ side where Li-ion is filling the vacancies. Such interaction-free treatment already gives a good description if an interpolation is made using the criterion of a positive chemical capacitance. Nonetheless we show how non-idealities such as ion-electron, ion-ion and electron-electron interactions can be introduced to obtain a mechanistic understanding in addition to just a pragmatic description. We intend to apply this concept also to compounds featuring a multi-electron reaction mechanism (e.g., Na$_3$V$_2$(PO$_4$)$_2$F$_3$ or Na$_3$V$_2$(PO$_4$)$_2$FO$_2$) where e.g. in the nano-crystalline state the full storage range can be observed.

References

1.	McKinnon, W. R.; Haering, R. R., Physical Mechanisms of Intercalation. In Modern Aspects of Electrochemistry: No. 15, White, R. E.; Bockris, J. O. M.; Conway, B. E., Eds. Springer US: Boston, MA, 1983; pp 235-304.

2.	Maier, J. Angewandte Chemie International Edition 2013, 52, (19), 4998-5026.

3.	Julien, C. M., Lithium intercalated compounds: Charge transfer and related properties. Materials Science and Engineering: R: Reports 2003, 40 (2), 47-102.

4.	Statistical thermodynamics of crystals containing point defects. In Atomic Transport in Solids, Lidiard, A. B.; Allnatt, A. R., Eds. Cambridge University Press: Cambridge, 1993; pp 92-160.

5.	Hainovsky, N.; Maier, J. Physical Review B 1995, 51, (22), 15789-15797.

6.	Kirkwood, J. G. The Journal of Chemical Physics 1946, 14, (3), 180-201.

7.	Mayer, J. E. The Journal of Chemical Physics 1950, 18, (11), 1426-1436.

8.	Allnatt, A. R.; Cohen, M. H. The Journal of Chemical Physics 1964, 40, (7), 1871-1890.

9.	Nitta, N.; Wu, F.; Lee, J. T.; Yushin, G., Li-ion battery materials: present and future. Materials Today 2015, 18 (5), 252-264.

10.	Yamada, A.; Koizumi, H.; Nishimura, S.-i.; Sonoyama, N.; Kanno, R.; Yonemura, M.; Nakamura, T.; Kobayashi, Y., Room-temperature miscibility gap in $Li_xFePO_4$. Nature Materials 2006, 5 (5), 357-360.

11. Maier, J.; Amin, R., Defect Chemistry of LiFePO4. Journal of The Electrochemical Society 2008, 155 (4), A339.

12. Amin, R.; Lin, C.; Maier, J., Aluminium-doped LiFePO4 single crystals Part II. Ionic conductivity, diffusivity and defect model. Physical Chemistry Chemical Physics 2008, 10 (24), 3524-3529.

13. Wagemaker, M.; Mulder, F. M., Properties and Promises of Nanosized Insertion Materials for Li-Ion Batteries. Accounts of Chemical Research 2013, 46 (5), 1206-1215.

14. Sanfeld, A., Thermodynamics of Surfaces. In Physical Chemistry: An Advanced Treatise, Jost, W., Ed. Academic Press: 1971; pp 245-291.

15. Maier, J., Chemical resistance and chemical capacitance. Zeitschrift für Naturforschung B 2020, 75 (1-2), 15-22.

16. Schubert, E. F., Semiconductor statistics. In Doping in III-V Semiconductors, Schubert, E. F., Ed. Cambridge University Press: Cambridge, 1993; pp 78-136.

17. Ion-Ion Interactions. In Modern Electrochemistry 1: Ionics, Bockris, J. O. M.; Reddy, A. K. N., Eds. Springer US: Boston, MA, 1998; pp 225-359.

18. Altland, A.; Simons, B., Second quantization. In Condensed Matter Field Theory, Cambridge University Press: Cambridge, 2010; pp. 39-94.

19. Lu, J.; Oyama, G.; Nishimura, S.-i.; Yamada, A. Chemistry of Materials 2016, 28, (4), 1101-1106.

20. Malik, R.; Zhou, F.; Ceder, G. Nature Materials 2011, 10, (8), 587-590.

Appendices

1. Voltage vs. $x_{Li}$ relationship under dilute conditions.

In a Li-ion battery cell with Li insertion material (M) as the working electrode and metallic Li as the reference electrode, the cell voltage is given by

$$V = -\frac{[\mu_{Li}(M) - \mu_{Li}(Li)]}{F}$$

For simplicity, let

$$\mu_{Li}(M) = \mu_{Li}$$

The Li chemical potential in M is also given by

$$\mu_{Li} = \mu_{Li}° + RT \ln a_{Li}$$

For very small Li contents (a few Li interstitials) in the fully neutral case

$$a_{Li} = x_{Li}$$

Therefore

$$-\frac{V \cdot F}{RT} = \ln x_{Li} + \text{const.}$$

For the fully dissociated case

$$\mu_{Li} = \mu_{Li_i^{\cdot}} + \mu_{e'} = \mu_{Li_i^{\cdot}}° + RT \ln a_{Li_i^{\cdot}} + \mu_{e'}° + RT \ln a_{e'} = RT \ln(a_{Li_i^{\cdot}} \cdot a_{e'}) + \text{const.}$$

Again for very small Li contents

$$a_{Li_i^{\cdot}} = a_{e'} = x_{Li}$$

Therefore

$$-\frac{V \cdot F}{RT} = \ln{(x_{Li})^2} + \text{const.} = 2\ln{x_{Li}} + \text{const.}$$

For very large Li contents (a few Li vacancies), the results are obtained analogously. For the fully neutral case

$$-\frac{V \cdot F}{RT} = -\ln{(1\text{-}x_{Li})} + \text{const.}$$

For the fully dissociated case

$$-\frac{V \cdot F}{RT} = -\ln{(1\text{-}x_{Li})^2} + \text{const.} = -2\ln{(1\text{-}x_{Li})} + \text{const.}$$

2. Criterium for phase stability.

For stability criteria to apply the phase must be locally stable, and does not need to be globally stable. Nonetheless, the application to the nano-crystalline state deserves a more detailed consideration. Here it suffices to state that we can safely apply them. Using the thermodynamic potential ($\Gamma$) considered in Ref. 15, all of these capacitances can be obtained by the second derivative of $\Gamma$ with respect to the intensive parameters (temperature, pressure, chemical potential of Li, etc.). The second derivatives of the $\Gamma$-function defined in Ref. 15 with respect to the intensive parameters give the characteristic capacitances: thermal capacitance, mechanical capacitance (compressibility) and chemical capacitance (variation of particle number with respect to chemical potential). In textbooks of physical chemistry (Ref. 14) it is shown that in order for a phase to be stable, all these capacitances need to be positive. This corresponds to a convex shape of the free energy curve. This stability means local stability and does not assume the absence of more favorable phases. The shape of the respective function becomes however critical in the spinodal regime of a mixed phase. For a nano-crystalline arrangement which is stabilized, e.g. by capillary effects or similar effects that influence the chemical potential, local stability should still be fulfilled. Problems may arise in the spinodal zone where the costly formation of interfaces prohibit an unmixing. However, for the present $Li_xFePO_4$ system, Gibbs-energy of $Li_xFePO_4$ for x from 0 to 1 looks rather monotonic ignoring very small variations (Ref. 20), and our criterion of phase stability should safely apply also for the nano-crystalline state, at least on the level of our approximation.

3. Saturation effects due to finite sites.

For Li contents that are not small, even ignoring defect interactions, Li activities instead of Li concentrations must be used because only finite sites are available for Li insertion. In the fully neutral case, Fermi-Dirac type of statistics gives

$$a_{Li} = \frac{x_{Li}}{1 - x_{Li}}$$

For the fully dissociated case

$$a_{Li_i} = \frac{[Li_i]}{1 - [Li_i]} = \frac{x_{Li}}{1 - x_{Li}}$$

In the dissociated case, saturation effects for ions and electrons have to be considered separately. The major difference here is electronic saturation. Here the simplest case can be described with the strictly localized approximation (i.e., electronic sites are saturated exactly the same way as ionic sites), namely

$$\text{Fe}^{3+} + \text{e}^- \rightleftharpoons \text{Fe}^{2+}$$

and

$$a_{e'} = \frac{[\text{Fe}^{2+}]}{1-[\text{Fe}^{3+}]} = \frac{[\text{Li}_i]}{1-[\text{Li}_i]} = a_{\text{Li}_i}$$

Figure A1 below shows a more drastic change of Li chemical potential changes in the dissociated model. The neutral model clearly falls short in not taking account of electronic defects.

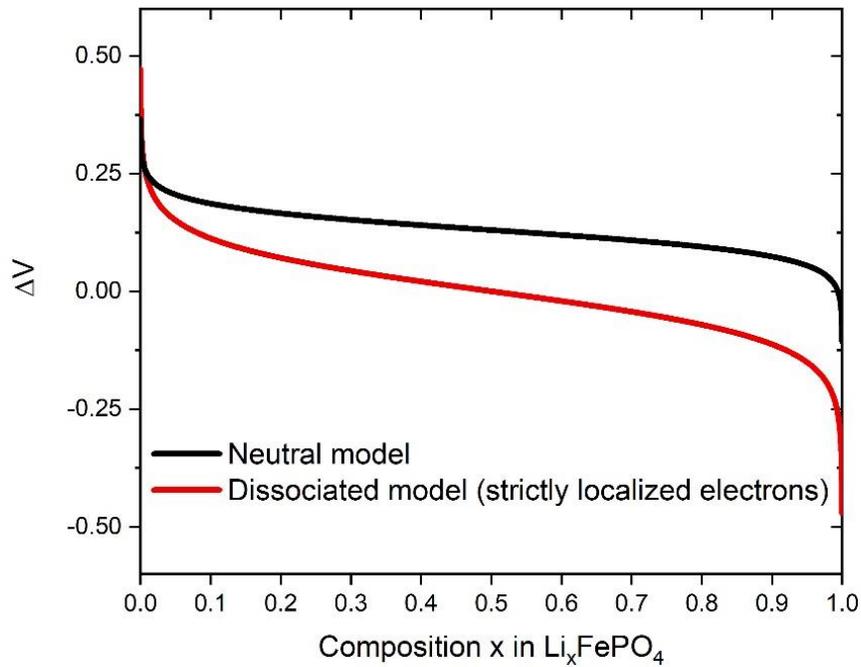

Fig. A1: Voltage change due to the saturation effect on both ionic and electronic sites.

4. Electron activities when filling the band.

If Fe-orbitals overlap as it is the case for $Li_xFePO_4$ quantum mechanics comes into play and the band structure needs to be considered.

By mass action law and considering saturation of sites

$$a_{Li} = a_{Li_i} \cdot a_{e'} = \frac{[Li_i]}{1 - [Li_i]} \cdot [[e'] \cdot f(e')]$$

where $f(e')$ is the activity coefficient of electrons given by Fermi-Dirac statistics. Unlike above now the energy of the level sensitively depends on occupation.

For fully dissociated case

$$x_{Li} = [Li_i] = [e']$$

Also

$$[e'] = \frac{n_{e'}}{N_A} \cdot V_m = \frac{n_{e'}}{N_C} \cdot \left(\frac{N_C \cdot V_m}{N_A}\right)$$

where $n_{e'}$ is the electron concentration, $N_A$ is the Avogadro constant, $V_m$ is the molar volume and $N_C$ is the effective density of states for the band to be filled.

Denote

$$n = \frac{N_A}{N_C \cdot V_m}$$

Then

$$\frac{n_{e'}}{N_C} = n \cdot x_{Li}$$

where n is the fitting parameter in the maintext.

A realistic value of n could be, for example,

$$n = \frac{N_A}{N_C \cdot V_m} = \frac{6.02 \times 10^{23}}{2.80 \times 10^{19} \times \frac{158}{3.68}} = 500$$

where the value of effective density of states is quoted from silicon and that of molar volume is calculated based on lithium iron phosphate.

Overall,

$$V = -\frac{\mu_{Li_i} + \mu_{e'}}{F} + \text{const.} = -\frac{RT}{F} \cdot \ln(\frac{[Li_i]}{1 - [Li_i]}) - \frac{RT}{F} \cdot \eta_F + \text{const.}$$

where

$$\eta_F = \frac{\mu_{e'}}{RT} + \text{const.} = \frac{\ln\frac{n_{e'}}{N_c}}{1 - \frac{n_{e'}}{N_c}} + \left(\frac{3}{4}\sqrt{\pi}\frac{n_{e'}}{N_c}\right)^{\frac{2}{3}} + \frac{\frac{3}{2}\sqrt{\pi}\frac{n_{e'}}{N_c}}{\left(3 + \frac{3}{4}\sqrt{\pi}\frac{n_{e'}}{N_c}\right)^2}$$

according to Nilsson's equation approximating the exact Fermi-Dirac integral (see Ref. 16).

Figure A2 below shows the effect of n, which is solely a result of filling available states of electrons, on the chemical potential of Li (hence the cell voltage).

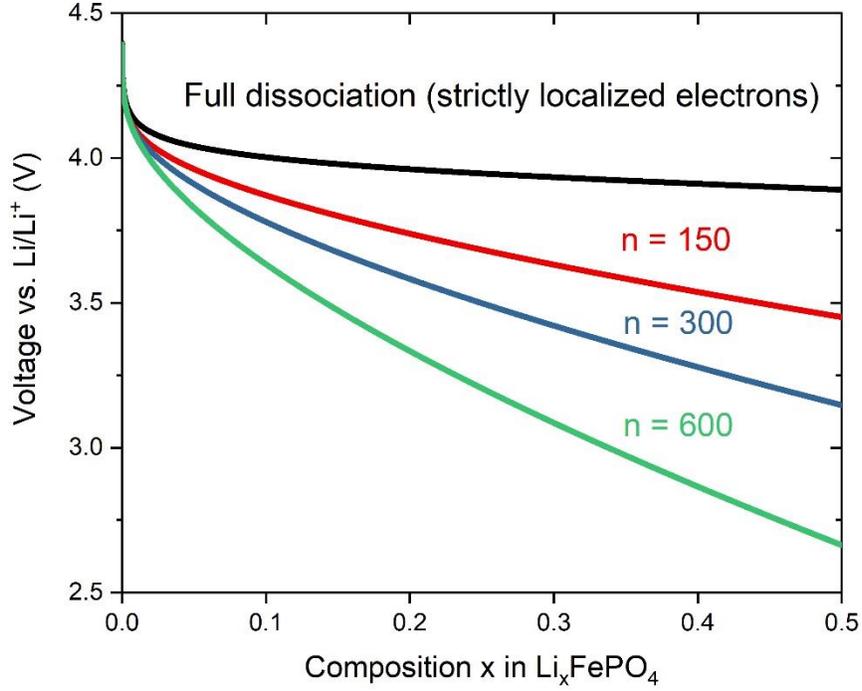

Fig. A2: Voltage change as a function of fitting parameter n.

5. Considering ion-electron interactions via association.

Here we consider

$$\text{Li}_i^{\cdot} + e' \rightleftharpoons \text{Li}_i^x$$

By mass action law and considering saturation of sites

$$a_{\text{Li}_i^x} = K_a a_{\text{Li}_i^{\cdot}} \cdot a_{e'} = K_a \frac{[\text{Li}_i^{\cdot}]}{1 - [\text{Li}_i^{\cdot}] - [\text{Li}_i^x]} \cdot \{[e'] \cdot f(e')\}$$

Therefore

$$K_a = \frac{\dfrac{[\text{Li}_i^x]}{1 - [\text{Li}_i^{\cdot}] - [\text{Li}_i^x]}}{\dfrac{[\text{Li}_i^{\cdot}]}{1 - [\text{Li}_i^{\cdot}] - [\text{Li}_i^x]} \cdot \{[e'] \cdot f(e')\}} = \frac{[\text{Li}_i^x]}{[\text{Li}_i^{\cdot}] \cdot \{[e'] \cdot f(e')\}}$$

where $K_a$ is the association constant and also the fitting parameter mentioned in the maintext.

Because

$$[Li_i'] = [e']$$

$$x_{Li} = [Li_i'] + [Li_i^x]$$

Hence

$$x_{Li} - [Li_i'] = [Li_i^x] = [Li_i'] \cdot \{[e'] \cdot f(e')\} \cdot K_a$$

$$\ln(x_{Li} - [Li_i']) = \ln([Li_i']) + \ln\{[e'] \cdot f(e')\} + \ln(K_a)$$

with

$$\ln\{[e'] \cdot f(e')\} = \eta_F = \frac{\ln\frac{n_{e'}}{N_c}}{1 - \frac{n_{e'}}{N_c}} + \left(\frac{3}{4}\sqrt{\pi}\frac{n_{e'}}{N_c}\right)^{\frac{2}{3}} + \frac{\frac{3}{2}\sqrt{\pi}\frac{n_{e'}}{N_c}}{\left(3 + \frac{3}{4}\sqrt{\pi}\frac{n_{e'}}{N_c}\right)^2}$$

Therefore, under a fixed n, the above non-linear equation can be solved numerically to obtain the value of $[Li_i']$ for each $x_{Li}$ at varying values of $K_a$.

Figures A3 and A4 below show how the degree of association and cell voltage are affected by association constant at a given n.

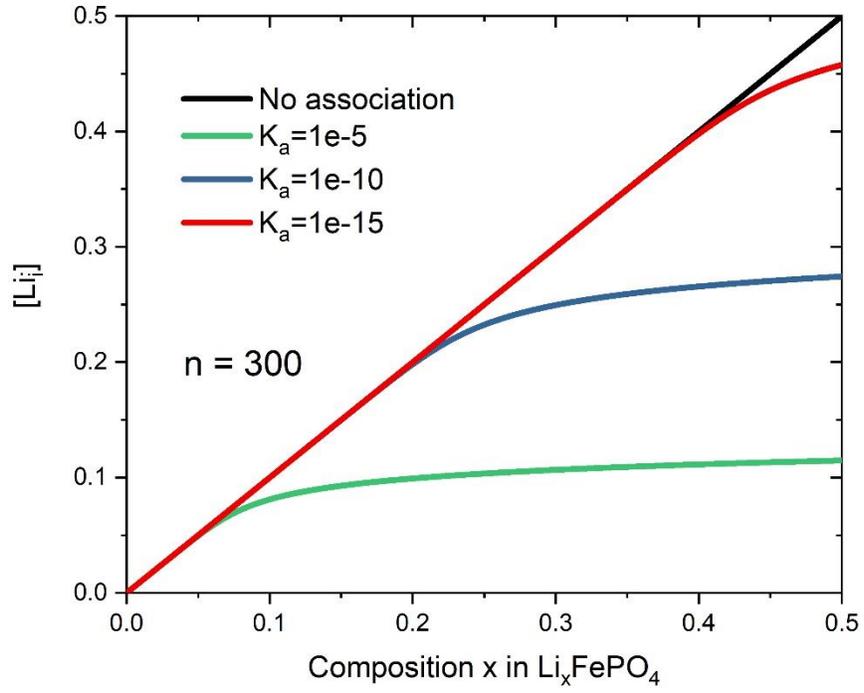

Fig. A3: Dissociated Li vs. inserted Li as a function of association constant $K_a$.

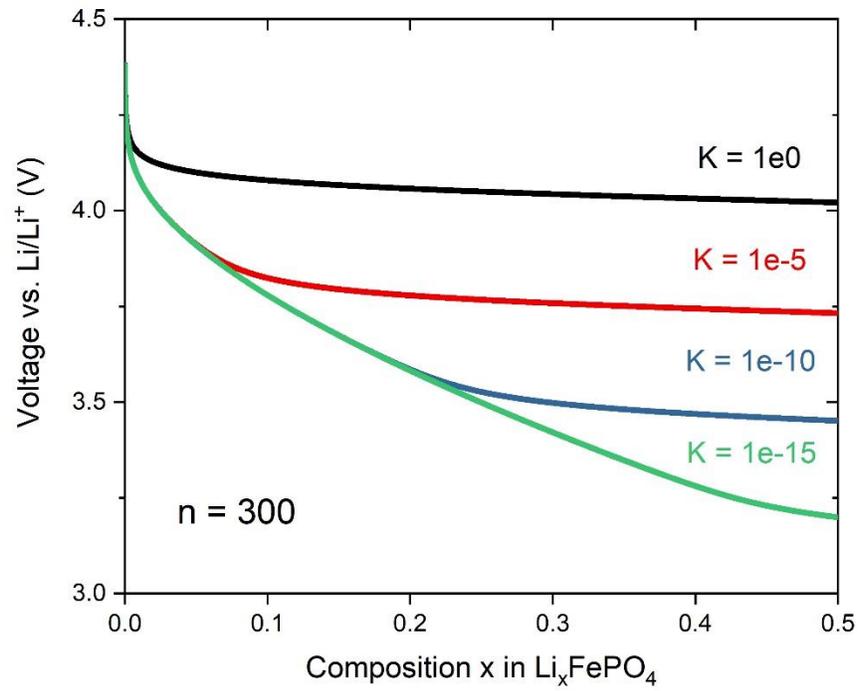

Fig. A4: Effect of the association constant $K_a$ on voltage change.

6. Overall fitting from both sides.

A larger range of satisfactory fitting can be obtained when both $n$ and $K_a$ are varied. Different sets of $n$ and $K_a$ can be attempted from the two ends. The best fitting value of $n$ is first determined, then followed by fitting to find the best $K_a$. Figures A5 and A6 below show the whole procedure.

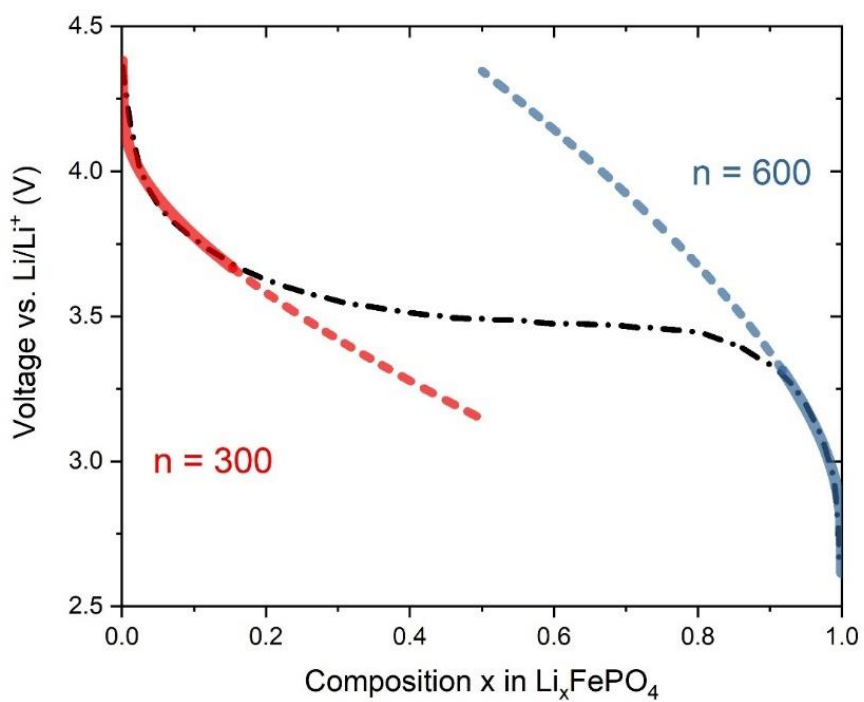

Fig. A5: Obtaining the best n.

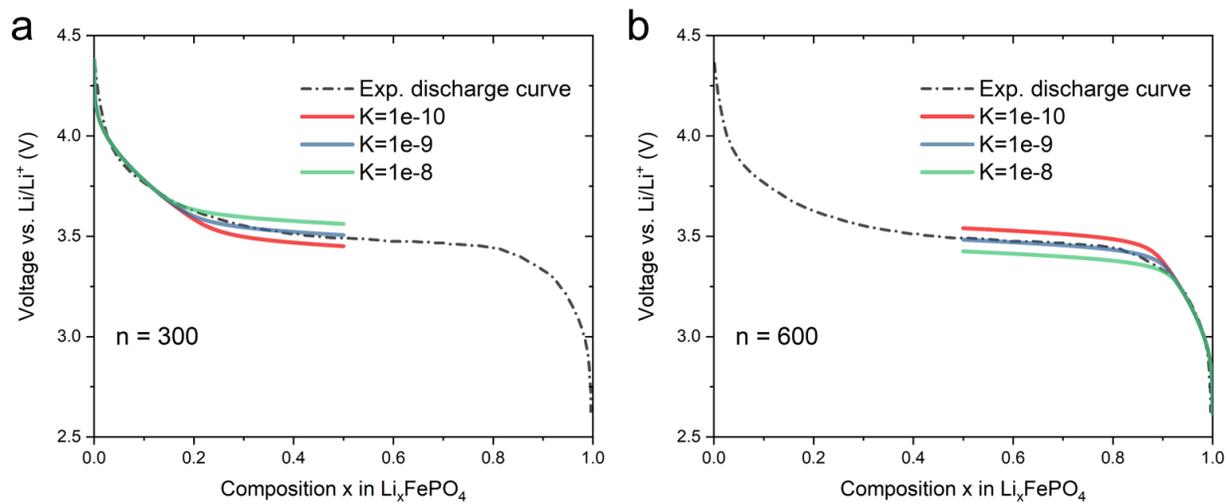

Fig. A6: Obtaining the best $K_a$ after having determined the best n.